\documentclass[12pt]{article}
\usepackage[left=3cm,top=3cm,right=3cm,bottom=2cm]{geometry}
\usepackage[]{natbib}      
\usepackage{url}
\setcitestyle{semicolon}
\usepackage{dcolumn}     
\usepackage{multirow}    
\usepackage{rotating}    
\usepackage{graphicx}    
\usepackage{bm}          
\usepackage{amssymb,amsmath}
\usepackage{eso-pic}
\makeatletter
\AddToShipoutPicture*{%
            \setlength{\@tempdimb}{.12\paperwidth}%
            \setlength{\@tempdimc}{.53\paperheight}%
            \setlength{\unitlength}{1pt}%
            \put(\strip@pt\@tempdimb,\strip@pt\@tempdimc){%
        \includegraphics[angle=30,width=6.5in]{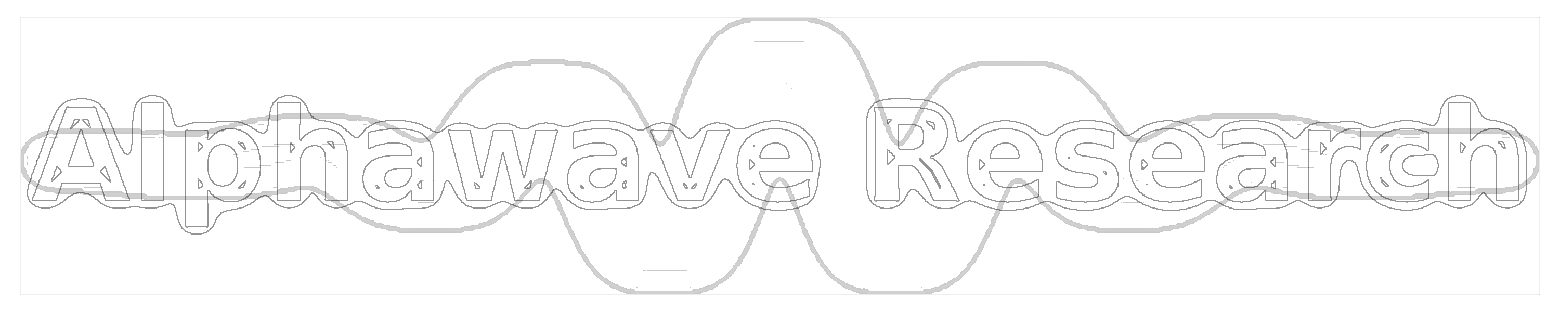}
            }%
}
\makeatother

\author{Robert W. Johnson \\
\small Alphawave Research\\[-0.8ex]
\small Atlanta, GA, USA\\
\small \texttt{robjohnson@alphawaveresearch.com}\\}
\title{Stationary axial plasma equilibrium in light of the magnetic polarization force}
\date{April 13, 2010\\ 
\small keywords: plasma equilibrium; plasma magnetization; polarization force}

\newcommand{\beq}{\begin{equation}}
\newcommand{\eeq}{\end{equation}}
\newcommand{\bea}{\begin{eqnarray}}
\newcommand{\eea}{\end{eqnarray}}
\newcommand{\mbf}[1]{\mathbf{#1}}
\newcommand{\mvec}[1]{\boldsymbol{#1}}

\newcommand{\thetahat}{\hat{\theta}}
\newcommand{\zhat}{\hat{z}}
\newcommand{\ptild}{\widetilde{p}}

\newcommand{\del}{\nabla}
\newcommand{\divr}{\nabla \cdot}
\newcommand{\curl}{\nabla \times}
\newcommand{\half}{\dfrac{1}{2}}
\newcommand{\oover}[1]{\dfrac{1}{#1}}
\newcommand{\ddr}[1]{\dfrac{\partial\, {#1}}{\partial r}}

\newcommand{\dext}{\mathrm{d}}
\newcommand{\cbr}[1]{{#1}}

\begin{document}
\maketitle

\begin{abstract}
The standard equilibrium equation for magnetized plasma is extended to account for the magnetic polarization force.  A factor of the pressure gradient arising from the magnetic decomposition of the Hall term survives the limit of vanishing magnetization and is canceled by a contribution from the magnetic polarization force, such that the proposed equation reduces to the standard form in the free current limit.  Comparison of the solutions for an axially symmetric plasma column indicates agreement with the free current limit for vanishing magnetization and variation in the field and current profiles as the degree of magnetization increases.
\end{abstract}


\section{Introduction}
\label{intro}
The full effect of plasma magnetization remains poorly understood.  From early explorations~\citep{grad-58,roseclark,shafranov-66,nasatnd-3880,vlasov-68,hall-710} through modern developments~\citep{hazeltine-03,hazeltine-04}, we contend that an important effect has been neglected in the analysis of plasma equilibrium, the magnetic polarization force, despite its experimental applications in fusion~\citep{hayes-213}, magnetic fluids~\citep{zahn-144,rinaldi-2847,eldib-159}, biophysics~\citep{qi-132,ataka-ic0038,wang-877,tzir-077103,rama-297} and materials science~\citep{cantor-02,colli-58,takagi-842,kozuka-884,asai-r1,maki-066106,ma-2944,ono-608,keil-07,fangwei-024202}.  \cbr{Its nonlinear nature is often cited when neglecting plasma magnetization in the usual introductory equilibrium equation}~\citep{chen-84,dendybook-93,staceybook05}, which treats the combined bound and free current densities as a single entity; however, we find that the reciprocal relationship between $M$ and $H$ leads to simplifications, rather than complications, in the treatment of stationary equilibrium in a strongly magnetized plasma.  Note that it is a weak applied field $\mbf{H}$ that produces the strongest magnetization $\mbf{M}$ for a given pressure $p$.

Following a brief review of conventional plasma equilibrium, we will examine the nonlinear model for plasma diamagnetism and its relation to the magnetic polarization force.  Next we consider the stationary equilibrium equation with the inclusion of magnetization to the net force balance.  With restriction to an axially symmetric configuration, we develop a model in the rest frame of the plasma for a magnetically confined column supporting an axial current.  While there are several differences with the usual model for solar coronal loops [see~\citep{tsypin-35544} and references therein], most obviously the presence of gravity and the curvature of the column, we will neglect such effects in order to get at the root of the magnetization problem.  Application to axial experimental configurations~\citep{carter-9410,palmer-0609} is apparent.  Performing a numerical evaluation, we compare the solutions of the magnetically decomposed equation both with and without the magnetic polarization force with that obtained in the free current limit, where we find agreement only with its inclusion.

\section{Conventional plasma equilibrium}
\label{conventional}
There are many ways to view a collection of free electrons and ions immersed in a background electromagnetic potential.  The two most common approaches address either the kinetic equations or the fluid equations, and collisional transport theory is used to relate them~\citep{fitzpatrick-notes}.  The kinematically conserved quantities are mass, momentum, and energy, each related to an increasing velocity moment of the transport equation.  The conventional plasma fluid equations rely on the statement that ``all charges and currents are considered to be free'' rather than utilize the formalism of macroscopic electrodynamics, and of primary importance is the influence of the Lorentz force on an electrically neutral medium $\mbf{J} \times \mbf{B}$.  In its simplest form neglecting gravity, the stationary equilibrium net force balance equation (also known as the momentum equation) for the single fluid model, \beq \label{eqn:pJB}
\del p = \mbf{J} \times \mbf{B} \;,
\eeq represents the balance of the Lorentz force due to plasma currents and the collision force identified as the pressure gradient~\citep{woodsbook-04} in the absence of changes to the momentum through Newton's second law.  Much technology~\citep{Lao:1985mw} goes into the solution of that equation in various circumstances.  The geometric form of Equation~(\ref{eqn:pJB}) indicates that the current and magnetic field vectors point along isobaric surfaces; that is, $\mbf{J} \cdot \del p = 0$ and $\mbf{B} \cdot \del p = 0$ for $\mbf{J} \times \mbf{B} \neq 0$.  The use of a scalar pressure rather than a gyrotropic tensor is often considered adequate for the analysis of plasma devices~\citep{frc-pop-2006}, and the distinction will not be considered here.  One commonly works in the rest frame of the plasma so that the fluid velocity $\mbf{V}_f$ makes no appearance.

\cbr{One's mathematical model should reflect the nature and scale of the physical system considered.  The fluid model in this article} is most appropriate for the description of macroscopic systems of sufficient density and extent that equilibrium thermodynamics is applicable.  While cold plasmas are easily confined by material boundaries, the goal of hot plasma confinement is to keep the ions sitting still (on average) by forcing electrons through the gas while applying magnetic fields from external sources.  A plasma is often considered a perfect conductor, in that the more it conducts, the more it wants to conduct, as the resistance decreases when its temperature increases.  It should also be considered a perfect diamagnet, with a susceptibility on the order of unity at high pressures, as those free charges are given much latitude for gyration in response to an applied magnetic field.

Our approach reflects that of classical field theory, combining macroscopic electrodynamics with continuum fluid mechanics, and we summarize our notation and assumptions: fully ionized hydrogenic plasma of species $s \in \{e,i\}$ with total particle density $n \equiv n_e + n_i$, total pressure $p \equiv n\,T = \sum_s n_s T_s$, mass density $\rho_m \equiv n\,m = \sum_s n_s m_s$, free current density $\mbf{J}_f \equiv \sum_s n_s e_s \mbf{V}_s$, vanishing charge density $\rho_e \equiv \sum_s n_s e_s = 0$ and mass velocity $\rho_m \mbf{V}_f \equiv \sum_s n_s m_s \mbf{V}_s = 0$, and isotropic pressure $W^\perp_s = 2 W^\parallel_s = T_s$ where $T_s \leftarrow k_B T_s$.  Our units are SI, but we express thermal energy $T$ in eV.  Note the total particle density is twice the neutral plasma density $n = 2 n_0$.  Our treatment of magnetization most closely follows that by~\citet{griffiths-89,hazeltine-04}; and~\citet{kauf-9212}, except that we will be fully decomposing the equilibrium equation in terms of $\mbf{H}$ and $\mbf{M}$, a procedure not found in the existing plasma physics literature.

\section{Diamagnetism, $\beta$ limit, and the magnetic polarization force}
\label{diamagnetism}
The fluid dipole moment per unit volume for each species is taken as $\mbf{M}_s \equiv n_s \mvec{\mu}_s$ \cbr{for $\mvec{\mu}_s \equiv -(W^\perp_s / B_s^2) \mbf{B}_s$}, where the field $\mbf{B}_s$ felt by a single particle of species $s$ within the unit of volume is the net field less the particle's own contribution $\mbf{B}_s / \mu_0 \equiv \mbf{H}+\mbf{M}-\mvec{\mu}_s = \mbf{H}+\mbf{M}_k + \alpha_s \mbf{M}_s$, where $k \neq s$ and \cbr{$\alpha_s \equiv (n_s - 1)/n_s$} is a unitless factor.  For a sufficiently dense plasma, \cbr{$n_s \gg 1$ such that} $\alpha_s \rightarrow 1$ and $\mbf{B}_s \rightarrow \mbf{B}$.  The total magnetization of the neutral fluid is given by the net dipole density, which for $\ptild \equiv p / \mu_0$ and $\mbf{h} \equiv \mbf{H}/H$ may be written \cbr{as the total number of particles times their average dipole moment,} \bea 
\mbf{M} &\equiv& \sum_s \mbf{M}_s = - \sum_s \left( \ptild_s / \vert \mbf{H}+\mbf{M} \vert^2 \right) \left( \mbf{H}+\mbf{M} \right) \;, \\
&=& - \mbf{h} M  = - \mbf{h} \ptild / \left(H-M\right) = 2 n_0 (\mvec{\mu}_e + \mvec{\mu}_i )/2 \;,
\eea and has~\citep{rwj-jpp03} the physical solution $M/H = [1 - (1 - 4 \ptild / H^2)^{1/2}]/2$ as the plasma is diamagnetic~\citep{marshall-122367}.  Ultimately, the proper treatment of magnetization requires the use of quantum theory, in particular as to account for spin~\citep{halzenmartin}.  From the form of the solution for $M$ one can immediately read a limit on the ratio of kinetic to free magnetic pressure, $\beta_H \equiv 2 \ptild / H^2 \leq 1/2$ for a dense plasma. In terms of the net field $B$, we have $\beta_B \equiv 2 \ptild / (H-M)^2 = \beta_H / (1 - M/H)^2 \leq 2$, and the ratio $M/H$ is limited to $1/2$.  In the dilute fluid limit $n_0 \rightarrow 1/r_i^3$ such that $\alpha \equiv (n-1)/n = 1 - 1/2 n_0 \rightarrow 1/2$, we find $\mbf{M}_s \rightarrow \mvec{\mu}_s$ so that the limits $\beta_H \rightarrow M/H \rightarrow 1$ when $T_i = T_e$.  One must be careful to define the appropriate unit of volume for a \cbr{magnetized plasma, which should be on an order no smaller than the cube of the ion gyroradius $r_i$ as the description of $\mvec{M}$ in terms of a bound current density requires at least a single orbit within the volume unit; beyond that scale one must address by trajectory charges passing through the region.}

\begin{figure}[t]
\includegraphics[scale=.5]{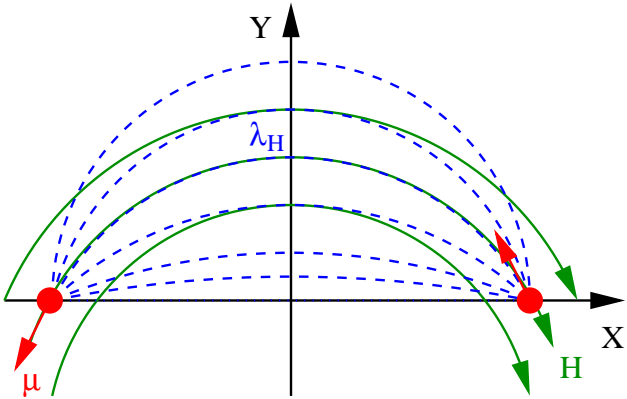}
\caption{Virtual trajectories (dashed) for the guiding center in a curved magnetic field (solid).  The magnitude of $\mvec{\mu}$ should be considered fixed and its direction tangential to the guiding center trajectory.  The path $\lambda_H$ along the field line defines a contour of $\mvec{\mu} \cdot \mbf{H}$, hence the force $\mu_0 \del (\mvec{\mu} \cdot \mbf{H})$ is everywhere perpendicular to $d \vec{\lambda}_H$ and the work done $\mu_0 \int_{\lambda_0}^{\lambda_1} d \vec{\lambda}_H \cdot \del (\mvec{\mu} \cdot \mbf{H})$ is zero.}
\label{fig:1}
\end{figure}

An infinitesimal dipole $\mvec{\mu}_s$ immersed in a field $\mbf{B}_s$ experiences a force $\mbf{f}_s = \del (\mvec{\mu}_s \cdot \mbf{B}_s)$ on the basis of the general theory of magnetized material~\citep{griffiths-89,aipdeskref,haus-mit}.  \Citet{feynmanlecs} gives a nice derivation in terms of virtual work.  This force felt by the guiding center is here generalized to the dipole density by $\mbf{F}_B = \del (\mbf{M} \cdot \mbf{B})$.  One may argue that the fluid should not exert a force on itself, and so we will also consider $\mbf{F}_H = \mu_0 \del (\mbf{M} \cdot \mbf{H})$, with either denoted by $\mbf{F}_M$.  We will not yet attempt a kinetic justification for the presence of this term among the moments of the Vlasov equation, suggesting that it should be instated at the level of the collisional Boltzmann equation, but rather will explore the consequences on the equilibrium equation of its inclusion, appealing to the mathematically well defined limiting process used in the reductions demonstrated below.  The effect of the magnetic force $\mbf{f}_s$ on an individual dipole may be understood heuristically as the force of constraint keeping the guiding center on track with its magnetic field line in the thin flux tube picture~\citep{ferriz-03,somov-06}, and the corresponding torque $\mbf{t}_s = \mvec{\mu}_s \times \mbf{B}_s$ keeps the guiding center aligned, as in Figure~\ref{fig:1}.  The established formalism does not care {\it how} $\mvec{\mu}_s$ is to appear, only that it {\it has} appeared, and the analogous effects for ferromagnetic materials should be familiar to anyone who has ever handled permanent magnets \cbr{in a laboratory setting or otherwise}.  Thus, $\mbf{J} \times \mbf{B} + \mbf{F}_M$ is our generalization of the macroscopic force densities~\citep{melcher-81,rosen-82} given by Lorentz and Kelvin, $\mbf{F}_\mathrm{LK} = \mu_0 \mbf{J} \times \mbf{H} + \mu_0 \mbf{M} \cdot \del \mbf{H}$, and by Korteweg and Helmholtz, $\mbf{F}_\mathrm{KH} = \mbf{J} \times \mbf{B} - \mbf{H} \cdot \mbf{H} \del \mu / 2$.  One may argue with our form for the magnetic polarization force on two grounds, as for finite gyroradius the expression for the force per particle $\mbf{f}_s$ is approximate and as we have chosen to differentiate the dipole density.  However, our choice leads to convenient simplifications and may be considered a first step towards a full accounting of magnetization effects on plasma equilibrium.

\section{Magnetized equilibrium equation}
\label{equilibrium}
The equilibrium net force balance equation we consider is given by \bea
\label{eqn:eq3} \del p = \mbf{J}_p \times \mbf{B} + \mbf{F}_M \; ,
\eea and the magnetic field is determined by $\divr \mbf{B} = 0$ and $\curl \mbf{B}_X = \mu_0 \mbf{J}_X$, where the subscript $X$ appearing in Ampere's law identifies the appropriate current source for each component of the magnetic field $\mbf{B}/ \mu_0 \equiv \mbf{H}+\mbf{M}$.  (Equilibrium in the presence of gravity requires the addition of the gravitational force $\mbf{F}_g \equiv \rho_m \mbf{g}$ to the RHS of the force balance equation above~\citep{yoshikawa-513,throum-0104072}.)  We require of our proposed equilibrium equation that it reduce to the standard form $\del p  = \mu_0 (\curl \mbf{H}) \times \mbf{H}$ in the limit of vanishing magnetization, which we identify as the free current limit $M \ll H$ such that $J_b \ll J_f$.  The total current density is the sum of the external current density and the plasma current density $\mbf{J}=\mbf{J}_{ext}+\mbf{J}_p$, where the plasma current is the sum of the free and bound currents $\mbf{J}_p \equiv \mbf{J}_f + \mbf{J}_b = \curl (\mbf{H}-\mbf{H}_{ext}) + \curl \mbf{M}$, giving us the magnetized equilibrium equation \bea \label{eqn:equil1}
\del p = \mu_0 \left[\curl (\mbf{H}+\mbf{M}) \right] \times (\mbf{H}+\mbf{M}) + \mbf{F}_M \; ,
\eea and we will show that the magnetic polarization force cancels a contribution from the magnetic decomposition of the Hall term which survives the limit of vanishing magnetization $\ptild \ll H^2$, noting that the curl of $\mbf{H}_{ext}$ is zero within the plasma as well as that the bound current, defined as the curl of the magnetization, remains divergence-free regardless of the geometry and includes the effects of the pressure gradient driven diamagnetic current and the curvature and $\del B$ drift currents $\curl \mbf{M} = - \curl (p / B^2) \mbf{B}$.  Note that bound charges and bound currents are physically distinct entities; the former are governed by Gauss's law and associated with the presence of a binding energy between unlike charges while the latter are governed by the Maxwell-Ampere equation and associated with the presence of charges undergoing microscopic circulation.  The gyromotion of plasma particles constitutes just such a microscopic circulation and is amenable to treatment as the magnetization of a fluid.

Using the vector identity $\left(\curl \mbf{B} \right) \times \mbf{B} = \left(\mbf{B} \cdot \del \right) \mbf{B} - \del ( \mbf{B} \cdot \mbf{B} ) / 2$ to decompose the Hall term into curvature and gradient contributions, we consider first the limit of the curvature term.  Denoting $\lim_{\ptild \ll H^2}$ by $\Rightarrow$ we find $M/H \Rightarrow 0$ and \bea
(\mbf{B}/\mu_0 \cdot \del) \mbf{B}/\mu_0 &=& \left[ (1-M/H) \mbf{H} \cdot \del \right] (1-M/H) \mbf{H} \\
 &=& (1-M/H)^2 (\mbf{H} \cdot \del) \mbf{H} - (1-M/H) \mbf{H} \left[ \mbf{H} \cdot \del (M/H) \right] \\
 &\Rightarrow& (\mbf{H} \cdot \del) \mbf{H} \;.
\eea  The gradient term may be written \bea
- \half \del (B/\mu_0)^2 &=& - \half \del (H-M)^2 = - \half \del (H^2 + M^2 - 2 M H) \\
 &=& - \half \del H^2 \left[ 1 + (M/H)^2 \right] + \del M H \;,
\eea and the first term reduces to $-H \del H$.  The second term, however, survives the limit \bea
2 \del M H &=& \left[ H - (H^2 - 4 \ptild)^{1/2} \right] \del H + H \del \left[ H - (H^2 - 4 \ptild)^{1/2} \right] \\
 &=& \left[ 1 - (1 - 4 \ptild / H^2)^{1/2} \right] H \del H + H \del H - H \del (H^2 - 4 \ptild)^{1/2} \\
 &=& \left[ 2 - (1 - 4 \ptild / H^2)^{1/2} \right] H \del H - H (H^2 - 4 \ptild)^{-1/2} \del (H^2 - 4 \ptild)/2 \\
 &=& \left[ 2 - (1 - 4 \ptild / H^2)^{1/2} \right] H \del H - (1 - 4 \ptild / H^2)^{-1/2} (H \del H - 2 \del \ptild) \;,
\eea such that $\del M H \Rightarrow \del \ptild$.  The magnetic polarization force $\mbf{F}_M/\mu_0 \Rightarrow - \del \ptild$ serves to cancel this term so that the decomposed equation reduces to the standard form $\del \ptild = (\mbf{H} \cdot \del) \mbf{H} -H \del H$ in the free current limit.  Using $\mbf{F}_B = - \del p$, the curvature term is balanced by the gradient of the energy density $2 p + \mu_0 (H-M)^2/2  = p + \mu_0 (H^2 - M^2)/2$.  Note that a magnetic polarization force density of $n\, \mbf{f} = \sum_s n_s \mbf{f}_s$ would only enforce the reduction in a region of constant density where $\del p = n \del T$.

\section{Numerical evaluation}
\label{evaluation}
Restricting consideration to an axially symmetric plasma column with $\partial /\partial \theta \equiv \partial /\partial z \equiv 0$ embedded in a constant external magnetic field $\mbf{H}_{ext} = H_{ext}^0 \zhat$, the free current within the plasma is here supposed to be purely axial $\mbf{J}_f = J_f(r) \zhat$, giving rise to an polar magnetic field $\mbf{H}_f = H_\theta(r) \thetahat$.  The free field $\mbf{H}=\mbf{H}_{ext}+\mbf{H}_f$ satisfies $\divr \mbf{B} = 0$, leaving us to evaluate the net force balance.  Using the parameter $f_M \in \{0, 1\}$ to indicate the absence or presence of the magnetic polarization force, some algebra yields a differential equation for $H_\theta$, where using $\mbf{F}_B$ \bea
\ddr{} \left[ 2 (1 + 2 f_M) \ptild + H^2 + H (H^2 - 4 \ptild)^{1/2} \right] = - \oover{r} \left( \dfrac{H_\theta}{H} \right)^2 \left[ H + (H^2 - 4 \ptild)^{1/2} \right]^2 \;,
\eea
and using $\mbf{F}_H$ we have \bea
\ddr{} \left[ 2 \ptild + (1 + 2 f_M) H^2 + (1 - 2 f_M) H (H^2 - 4 \ptild)^{1/2} \right] = - \oover{r} \left( \dfrac{H_\theta}{H} \right)^2 \left[ H + (H^2 - 4 \ptild)^{1/2} \right]^2 \;,
\eea which are invariant under the transformation $H_\theta \rightarrow - H_\theta$ corresponding to the two possible orientations for the axial current.  The unmagnetized equilibrium equation is given by \bea \label{eqn:unmag}
\ddr{\ptild} + H_\theta \ddr{H_\theta} = - \dfrac{1}{r} H_\theta^2 \;,
\eea and our problem is defined as: given $p(r)$, find $H_\theta(r)$.

\begin{figure}[t]
\includegraphics[]{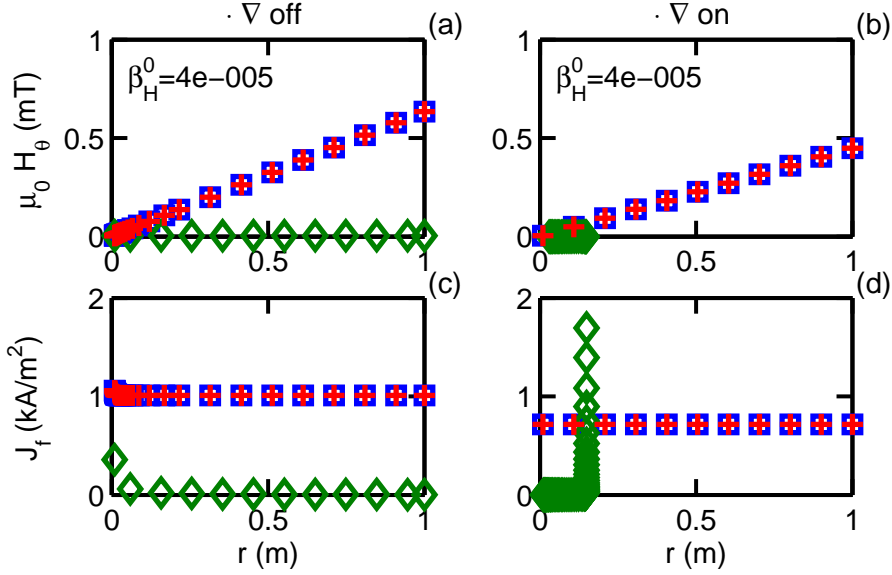}
\caption{Magnetized solutions with ($\Box$) and without ($\Diamond$) the magnetic polarization force $\mbf{F}_B$ compared to the free current model ($+$) for $p_0=10^{18}\mathrm{eV/m^3}$.  Curvature is neglected on the left and included on the right.  The magnetized model without the magnetic polarization force \cbr{does not agree with the free current limit}.}
\label{fig:2}
\end{figure}

The density of a solar coronal loop is on the order of $10^{14} \sim 10^{18} / {\rm m}^3$, and temperatures are found between 1eV and 1keV, with an polar field of a few mT and a column-aligned field up to several Tesla~\citep{tsypin-35544}.  Parametrizing the pressure of a plasma column of meter radius by $p(r)=p_0 (1 - r^a)$ with constant $a>0$ and using units of $\mathrm{eV/m^3}$, we consider the parabolic $a=2$ profile with a central pressure $p_0$ ranging from $10^{18}$ to $10^{22}$ in an external field of 100mT, ensuring that the central $\beta_H^0 < 1/2$.  We start by comparing the magnetized solutions both with ($f_M=1$) and without ($f_M=0$) the magnetic polarization force $\mbf{F}_B$ to that obtained by the free current model $H_\theta^2(r) = \ptild_0 r^2$ with $p_0=10^{18}$ in Figure~\ref{fig:2}.  Neglecting the curvature term ($\cdot \del$ off) introduces a factor $\sqrt{2}$ compared to its inclusion ($\cdot \del$ on) on the magnitude of the solution profile.  The solution for the magnetically decomposed equation without the magnetic polarization force in no way resembles the free current model's solution, which differs only slightly from the solution with the magnetic polarization force, and it diverges when the curvature term is included.

\begin{figure}[t]
\includegraphics[]{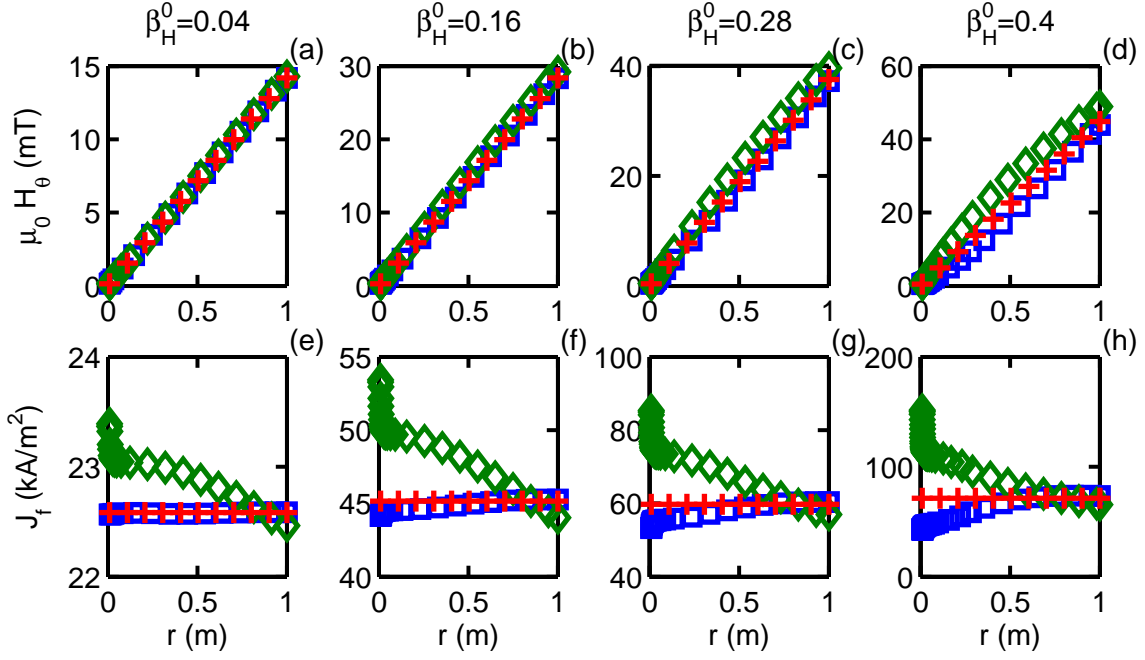}
\caption{Solutions for the free magnetic field (a)-(d) and the free current density (e)-(h) using either $\mbf{F}_B$ ($\Box$) or $\mbf{F}_H$ ($\Diamond$) compared to the free current model ($+$) for $p_0$ between $10^{21}$ and $10^{22}$.  The difference in the field between the models becomes apparent only at extremely high $\beta_H$.  The magnetic polarization force models differ in whether the central current should increase or decrease from the free current model's value.}
\label{fig:3}
\end{figure}

\begin{figure}[t]
\includegraphics[]{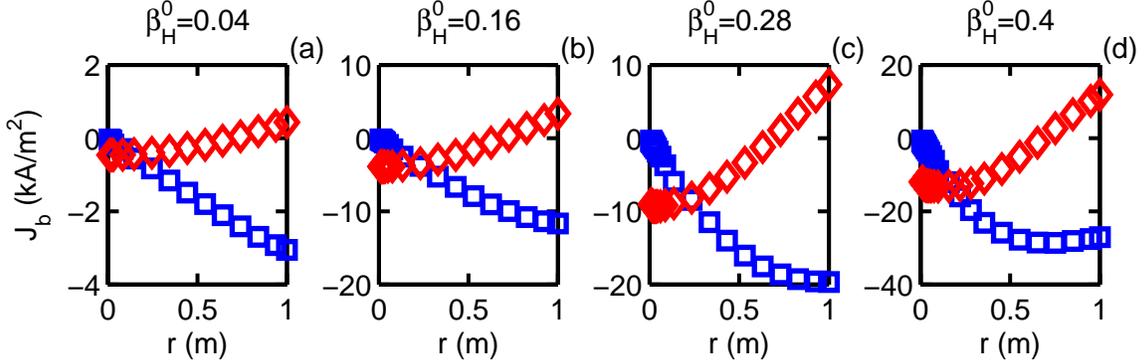}
\caption{Polar ($\Box$) and axial ($\Diamond$) bound current density solutions with $\mbf{F}_B$ for $p_0$ between $10^{21}$ and $10^{22}$.  Near the $\beta_H$ limit the bound axial current approaches $\pm20$\% the free axial current's value.}
\label{fig:4}
\end{figure}

Next we compare the solutions using $\mbf{F}_B$ and $\mbf{F}_H$ (including magnetic curvature) to the free current model for $p_0$ between $10^{21}$ and $10^{22}$ in Figure~\ref{fig:3}.  At moderate $\beta_H^0$ the solutions for $H_\theta$ are virtually identical, and only at extremely high $\beta_H^0$ do the magnetic polarization force models become distinguishable, where $\mbf{F}_B$ is more similar to the free current model.  The magnetic polarization force models differ in whether $J_f$ should be more or less than the free current model's value.   Finally, we compute the bound current $\mbf{J}_b = \mbf{H} \times \del (M/H) - (M/H) \curl \mbf{H}$ using the $\mbf{F}_B$ model for the same range of central pressures, shown in Figure~\ref{fig:4}.  The polar bound current generally exceeds the axial bound current, except near the core of the plasma column.  At the highest $\beta_H^0$ considered, the axial bound current ranges over $\pm$20\% of the axial free current's value, indicating that magnetization effects become pronounced as one approaches the $\beta_H$ limit.

\section{Conclusions}
\label{concl}
In conclusion, we have determined that our conjectured form for $\mbf{F}_M$ provides just the right cancellation required to take the magnetically decomposed $\mbf{H} + \mbf{M}$ equilibrium equation over to its usual form in the free current limit $M \ll H$ for either choice of magnetic polarization force model.  \cbr{The extra $\del p$ found upon taking the limit of the Hall term $\mvec{J} \times \mvec{B}$ stymies attempts to proceed with the reduction of the force balance to the free current limit without its inclusion.}  By providing the necessary balance, the magnetic polarization force $\mbf{F}_M$ allows for the extension of the standard equilibrium theory into the strongly magnetized regime as the $\beta_H$ limit is approached.

When compared to solutions of the free current model for an axially symmetric plasma column, we find considerable agreement between the magnetic polarization force models over a wide range of $\beta_H$.  Only in the limit $2 \ptild/ H^2 \rightarrow 1/2$ do the models become distinguishable, suggesting that experimental discrimination will be exceedingly difficult.  By isolating the free and bound currents, the magnetized equilibrium equation provides a more complete picture of what is happening inside the plasma.

\cbr{A model for plasma based on macroscopic electrodynamics is conceptually simple: Maxwell's theory of fields determines the potential from the source $A(J)$ through the geometric relation $\dext * \dext A = J$~\citep{ryder-qft,davis70}, and Newton's (or Einstein's) theory of kinematics determines the source from the potential $J(A)$ through the Ohm's law equation, which depends as well on the momentum $K(J,A)$.}  Ionized media is no less a material than any other mechanistic system, as its constituents must obey the physics dictated by their local environment and their interactions with other members.  While the field theory is more or less straightforward, the influence of those particle interactions on the kinematics is daunting to compute, and therein lies the difficulty and beauty of plasma physics.


\bibliographystyle{plainnat}

\end{document}